\documentclass[reprint,showpacs,preprintnumbers,amsmath,amssymb,longbibliography,aps,pra]{revtex4-1}
\usepackage[utf8]{inputenc}
\usepackage{mathrsfs}
\usepackage{amsmath}
\usepackage{bm}
\usepackage{amsfonts}
\usepackage{amssymb}
\usepackage{amsthm}
\usepackage{color}
\usepackage{graphicx}
\graphicspath{{./images/}}
\usepackage{geometry}
\usepackage{hyperref}
\usepackage{ulem}
\usepackage{epstopdf}
\usepackage{graphicx}
\hypersetup{
     unicode=false,
     pdftoolbar=true,
     pdfmenubar=true,
     pdffitwindow=false,     
     pdfstartview={FitH},    
     pdftitle={My title},    
     pdfauthor={Author},     
     pdfsubject={Subject},   
     pdfcreator={Creator},   
     pdfproducer={Producer}, 
     pdfkeywords={keyword1} {key2} {key3}, 
     pdfnewwindow=true,      
     colorlinks=true,       
     linkcolor=blue,          
     citecolor=blue,        
     filecolor=magenta,      
     urlcolor=blue           
}
\usepackage[toc,page]{appendix}

\begin{document}
\title{Off-axial focusing of spin-wave lens in the presence of Dzyaloshinskii-Moriya interaction}
\author{Weiwei Bao}
\author{Zhenyu Wang}
\author{Yunshan Cao}
\author{Peng Yan}
\email[Corresponding author: ]{yan@uestc.edu.cn}
\affiliation{School of Electronic Science and Engineering and State Key Laboratory of Electronic Thin Films and Integrated Devices, University of Electronic Science and Technology of China, Chengdu $610054$, China}
\begin{abstract}
We theoretically study the effect of Dzyaloshinskii-Moriya interaction (DMI) on the focusing of a spin-wave lens that is constructed by a circular interface between two magnetic films. We analytically derive the generalized Snell's law in the curved geometry and the position of the focal point which exhibits a peculiar off-axial focusing behavior. We uncover a strong dependence of the focal point on both the material parameters and the frequency of incident spin waves. Full micromagnetic simulations compare well with theoretical predictions. Our findings would be helpful to manipulate spin waves in chiral magnets and to design functional magnonic devices.
\end{abstract}
\maketitle
\section{Introduction}\label{sec1}
Spin waves (or magnons when quantized) are considered as potential data carriers for future information processing and logic operation, due to low energy consumption, (sub-)micron scale wavelength, and wide frequency range from GHz to THz \cite{Serga2010,Chumak2015}. Control of spin-wave propagation is crucial for application in magnonic devices.
Recently, there are growing interests to construct spin-wave lens by engineering the interface \cite{Toedt2016,Papp2018} or by modulating the refractive-index gradient \cite{Dzyapko2016,Whitehead2018,Vogel2019}. These spin-wave lens can focus a plane wave to a point with an enhanced amplitude, which is helpful for detecting weak spin-wave signal and the energy harvesting.

The Dzyaloshinskii-Moriya interaction (DMI) \cite{Dzyaloshinsky1958,Moriya1960} is the antisymmetric component of exchange coupling, which originates from the spin-orbit interaction in magnetic materials with broken inversion symmetry, either in bulk or at the interface. The DMI holds a chiral character and has led to a plethora of exotic phenomena such as nonreciprocal propagation of spin waves \cite{Zakeri2010,Cortes2013,Moon2013,Garcia2014}, magnon Hall effect \cite{Onose2010,Ideue2012}, negative refraction of spin waves \cite{Yu2016,Wang2018,Mulkers2018}, nonlinear three-magnon processes \cite{Wang2018,Zhang2018B}, and magnonic Goos-H\"{a}nchen effect \cite{Wang2019}, to name a few. Various chiral spin-wave devices have been proposed and designed, such as the spin-wave fiber \cite{Yu2016,Mulkers2018,Xing2016} and the spin-wave diode \cite{Lan2015}. The DMI effect on the spin-wave propagation thus opens an exciting window to observe rich physics and to realize functional devices.

It has been regarded as a well established notion that the focal point is on the axis of a geometrically symmetrical lens, when the wave beam propagates parallel to the lens axis \cite{Toedt2016,Papp2018,Whitehead2018,Vogel2019,Orloff2008,Mildner2011,Li2012}. In this work, we challenge this paradigm by theoretically investigating the DMI effect on the focusing of a semi-circular spin-wave lens (see Fig. \ref{fig1}). It is found that the DMI would cause a lateral shift of the focal point of spin waves, thus off the lens axis. Based on the generalized Snell's law, we analytically derive the coordinate of the focal point for spin-wave focusing. It shows that the induced lateral (horizontal) focal-point shift is an odd (even) function of the DMI parameter. Interestingly, we find that both the lateral and horizontal shifts increase with the increasing of the spin-wave frequency. However, external magnetic fields can suppress the shift in both directions. Our results are useful for understanding the spin-wave propagation in curved geometries and for designing functional magnonic devices in chiral magnets.

The paper is organized as follows. In Sec. \ref{sec2}, we present the theoretical model for the spin-wave propagation across the semi-circular interface, where the general Snell's law is derived to describe the wave scattering. We obtain the analytical formula of the focal-point coordinate. Section \ref{sec3} gives the results of micromagnetic simulations to verify theoretical predictions. Conclusions are drawn in Sec. \ref{sec4}.

\section{Analytical Model}\label{sec2}
\begin{figure}
  \centering
  \includegraphics[width=0.48\textwidth]{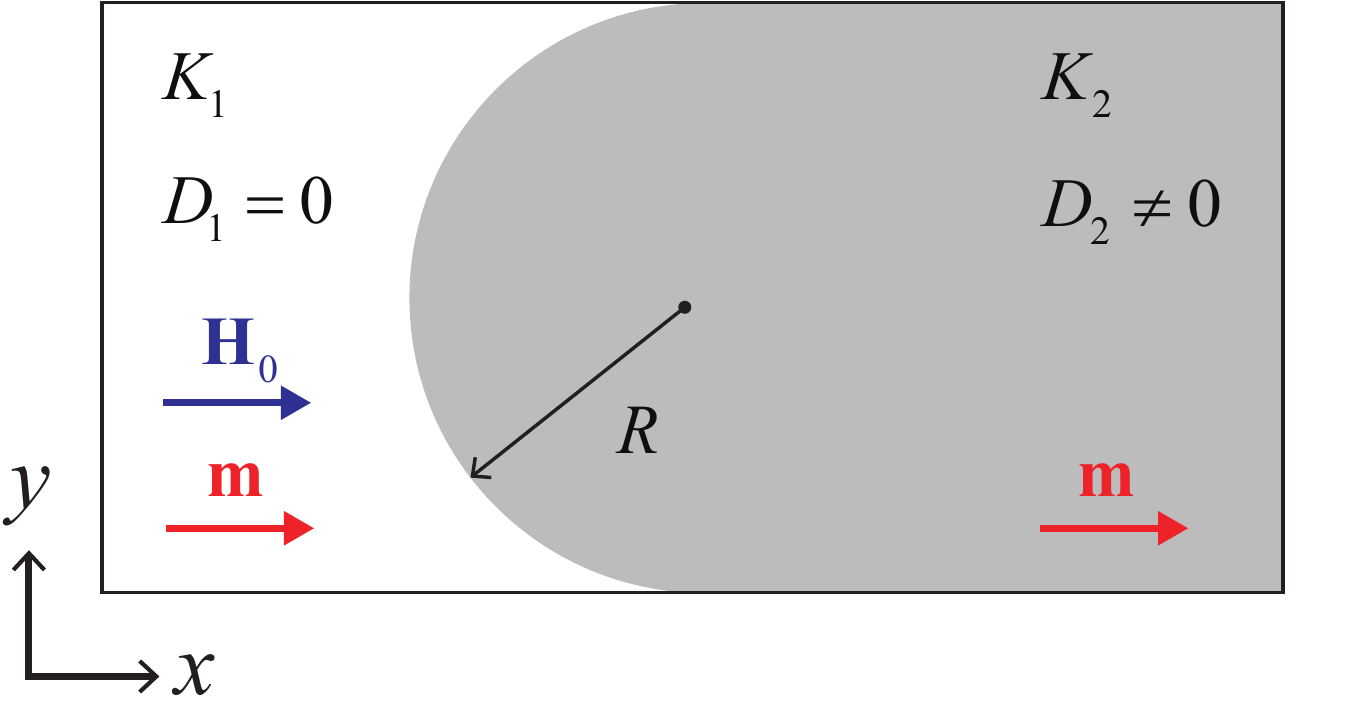}\\
  \caption{Schematic of a spin-wave lens with a semi-circular interface between two regions with different magnetic anisotropies $K_{1,2}$ and DMIs $D_{1,2}$. The radius of the semi-circular interface is $R$. The external magnetic field $\mathbf{H}_{0}$ is along $+\hat{x}$ direction to saturate the magnetization $\mathbf{m}$.}\label{fig1}
\end{figure}
\begin{figure}
  \centering
  \includegraphics[width=0.48\textwidth]{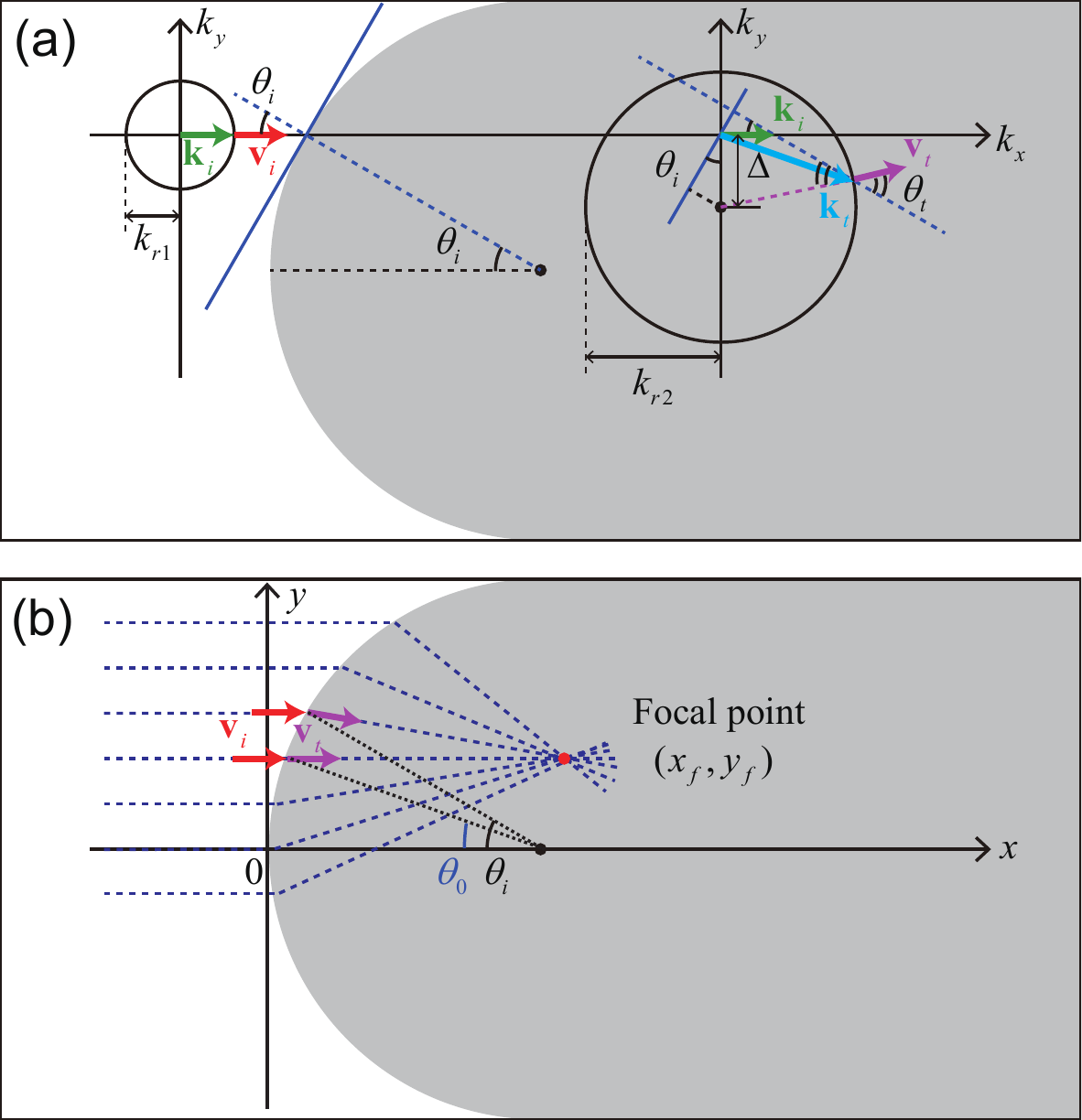}\\
  \caption{(a) Schematic plot of the generalized Snell's law for spin-wave scattering at the semi-circular interface. The solid and dashed blue lines represent the tangential and normal directions of the interface, respectively. $\mathbf{k}_{i,t}$ and $\mathbf{v}_{i,t}$ denote the wave vector and group velocity of the incident and refracted spin waves, with $\theta_{i}$ and $\theta_{t}$ being the incident and refracted angles, respectively. (b) Spin-wave focusing. Red and purple arrows label the group velocity of the incident and refracted spin waves, respectively. The dashed blue lines are the spin-wave rays while the red dot is the focal point $(x_{f},y_{f})$. $\theta_{0}$ corresponds to the special incident angle at which the group velocities of the incident and refracted spin waves are parallel with each other.}\label{fig2}
\end{figure}

We consider a spin-wave lens with a semi-circular interface between two ferromagnetic films with different anisotropies ($K_{1,2}$) and DMIs ($D_{1,2}$), as shown in Fig. \ref{fig1}. The magnetization dynamics is described by the Landau-Lifshitz Gilbert (LLG) equation
\begin{equation}\label{eq_llg}
  \frac{\partial\mathbf{m}}{\partial t}=-\gamma\mu_{0}\mathbf{m}\times\mathbf{H}_{\mathrm{eff}}+\alpha\mathbf{m}\times\frac{\partial\mathbf{m}}{\partial t},
\end{equation}
where $\mathbf{m}=\mathbf{M}/M_{s}$ is the unit magnetization vector with the saturated magnetization $M_{s}$, $\gamma=1.76\times 10^{11}$Hz/T is the gyromagnetic ratio, $\mu_{0}$ is the vacuum permeability, and $\alpha>0$ is the Gilbert damping constant. The effective field $\mathbf{H}_{\mathrm{eff}}$ comprises the exchange field, the DM field, the anisotropy field, the external magnetic field, and the dipolar field. The DMI considered here has the interfacial form \cite{Bogdanov2001}
\begin{equation}\label{eq_dmi}
  \mathbf{H}_{\mathrm{DM}}=\frac{2D}{\mu_{0}M_{s}}[\nabla m_{z}-(\nabla\cdot \mathbf{m})\hat{z}],
\end{equation}
where $D$ is the DMI constant. The magnetic anisotropy along the $x$-axis is assumed and the film is uniformly magnetized along the $+\hat{x}$ direction ($\mathbf{m}_{0}=+\hat{x}$). For simplicity, the dipolar interaction is approximated by the demagnetizing field $\mathbf{H}_{\mathrm{d}}=-M_{s}m_{z}\hat{z}$. For a small fluctuation of $\mathbf{m}$ around the equilibrium direction $\mathbf{m}_{0}$, we express the magnetization as $\mathbf{m}=m_{0}\hat{x}+m_{y}\hat{y}+m_{z}\hat{z}$ with $m_{0}\approx1$ and $|m_{y,z}|\ll1$.
Neglecting the damping term ($\alpha=0$), the spin-wave dispersion relation can be obtained by solving the linearized LLG equation \cite{Cortes2013,Moon2013}
\begin{equation}\label{eq_dispersion1}
  \omega(\mathbf{k})=\sqrt{(A^{\ast}\mathbf{k}^{2}+\omega_{H_{i}})(A^{\ast}\mathbf{k}^{2}+\omega_{H_{i}}+\omega_{m})}+D_{i}^{\ast}k_{y},
\end{equation}
where $A^{\ast}=2\gamma A/M_{s}$ with the exchange constant $A$, $D_{i}^{\ast}=2\gamma D_{i}/M_{s}$, $\omega_{H_{i}}=\gamma (2K_{i}/M_{s}+H_0)$ with the uniaxial anisotropy constant $K_{i}$, $\omega_{m}=\gamma\mu_{0}M_{s}$, and $\mathbf{k}=(k_{x},k_{y})$ is the wave vector of spin wave. Here $i=1,2$ represent parameters in the left domain (region 1) and right domain (region 2), respectively, as shown in Fig. \ref{fig1}. In order to facilitate the analysis of spin-wave propagation across the interface, we consider high-frequency spin waves, such that Eq. (\ref{eq_dispersion1}) can be simplified to
\begin{equation}\label{eq_dispersion2}
  \omega(\mathbf{k})=A^{\ast}\mathbf{k}^{2}+D_{i}^{\ast}k_{y}+\omega_{H_{i}}+\frac{\omega_{m}}{2}.
\end{equation}
Based on Eq. (\ref{eq_dispersion2}), we plot the isofrequency curves of the spin-wave propagation in two regions in $\mathbf{k}$ space, as shown in Fig. \ref{fig2}(a). In region 1, spin waves with a given frequency $\omega$ form a circle centered at the origin with the radius $k_{r1}=\sqrt{(\omega-\omega_{H_{1}}-\omega_{m}/2)/A^{\ast}}$. In region 2, the isofrequency circle is shifted by $\Delta=D_{2}^{\ast}/2A^{\ast}$ along the $-k_{y}$ axis and its radius becomes $k_{r2}=\sqrt{(\omega-\omega_{H_{2}}-\omega_{m}/2)/A^{\ast}+\Delta^{2}}$.

According to the continuity of the wave vector $\mathbf{k}$ tangent to the interface, we obtain the generalized Snell's law
\begin{equation}\label{eq_snell}
  k_{r1}\sin\theta_{i}+\Delta\cos\theta_{i}=k_{r2}\sin\theta_{t},
\end{equation}
where $\theta_{i}$ and $\theta_{t}$ are the incident and refracted angles with respect to the interface normal, see Fig. \ref{fig2}(a). We assume that spin waves are incident along $+\hat{x}$. So, through an arbitrary incident point $(x_{i},y_{i})=(R-R\cos\theta_{i},R\sin\theta_{i})$ on the semicircle, the equation of the refracted ray can be written as
\begin{equation}\label{eq_refracted}
  y_{t}=\tan(\theta_{t}-\theta_{i})[x_{t}-R(1-\cos\theta_{i})]+R\sin\theta_{i},
\end{equation}
where $(x_{t},y_{t})$ is the coordinate of an arbitrary point in the refracted beam. From Eqs. (\ref{eq_snell}) and (\ref{eq_refracted}), it is obvious to see that the intersection of refracted rays cannot be focused to a single point, which is the so-called spherical aberration caused by the circular shape of the interface. A perfect focusing only happens in the small angle limit, while the ideal shape of the interface to focus all spin waves can be constructed by some sophisticated methods \cite{Toedt2016,Papp2018}. In our case, the focal point is expected to be present on a refracted spin-wave branch that is parallel with the original incident beam carrying the incident angle
\begin{equation}\label{eq_tht0}
  \theta_{0}=\arctan\Big(\frac{\Delta}{k_{r2}-k_{r1}}\Big).
\end{equation}
The intersection point of the branch with other refracted rays is then given by
\begin{equation}\label{eq_coordinate}
  \begin{split}
    x &= \frac{R(\sin\theta_{0}-\sin\theta_{i})}{\tan(\theta_{t}-\theta_{i})}+R(1-\cos\theta_{i}), \\
    y &= R\sin\theta_{0}.
  \end{split}
\end{equation}
\begin{figure}
  \centering
  \includegraphics[width=0.48\textwidth]{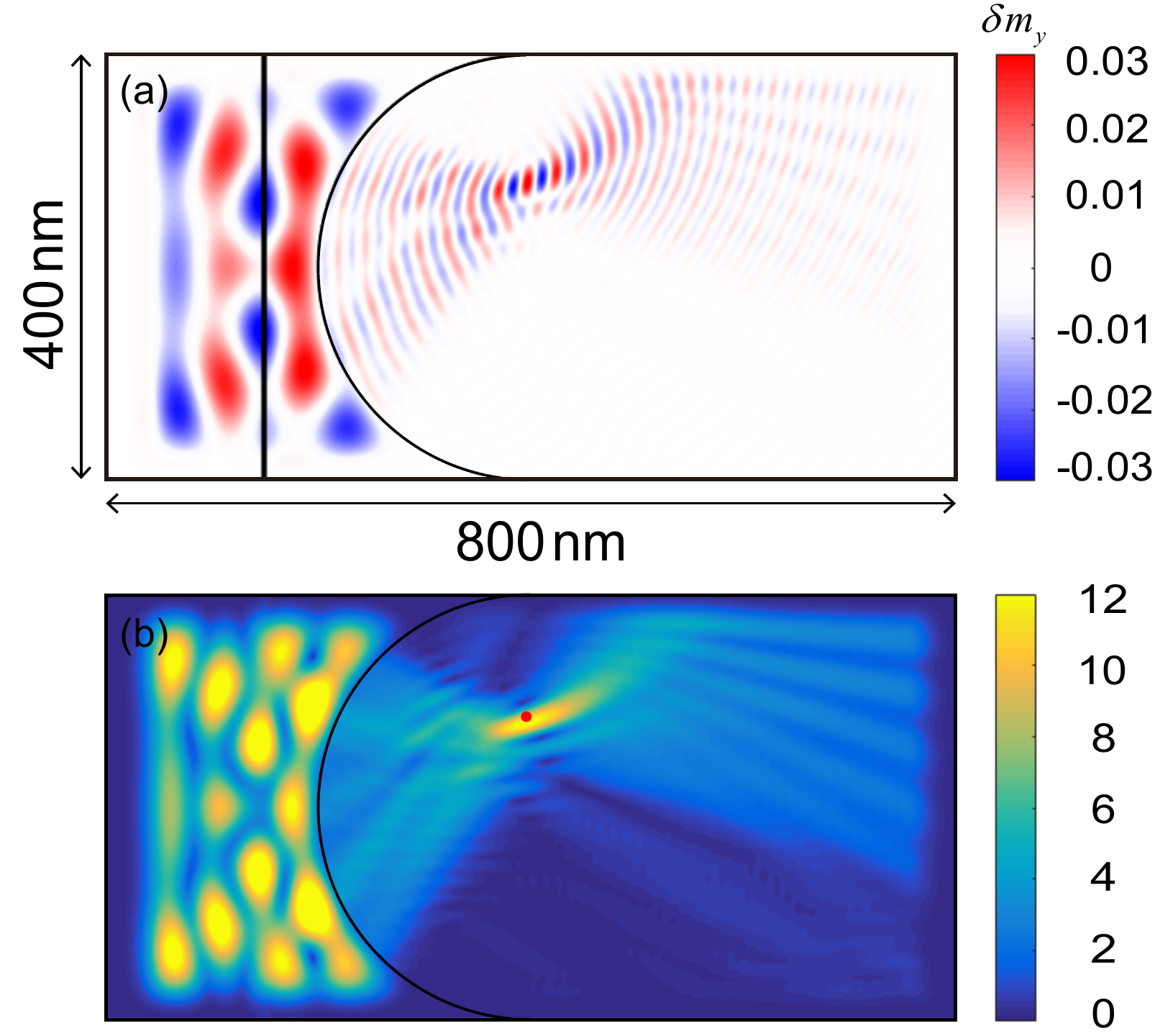}\\
  \caption{(a) Snapshot of micromagnetic simulation of the spin-wave focusing across the semi-circular interface with $\omega/2\pi=75$ GHz and $D_{2}=3$ $\mathrm{mJ/m^{2}}$. The black curve represents the interface of the spin-wave lens. The thick black line shows the source for spin-wave excitations. (b) Intensity map of spin waves in (a). The red point denotes the focal point obtained from the analytical formula (\ref{eq_fp}).}\label{fig3}
\end{figure}Considering the paraxial (or small-angle) approximation, we obtain the analytical formula of the focal-point coordinate
\begin{equation}\label{eq_fp}
  \begin{split}
    x_{f} &= \frac{k_{r2} }{k_{r2}-k_{r1}}R\cos^{3}\theta_{0}+R(1-\cos\theta_{0}), \\
    y_{f} &= R\sin\theta_{0},
  \end{split}
\end{equation}
by taking the limit $\theta_{i}\rightarrow\theta_{0}$ in Eq. (\ref{eq_coordinate}). Equation (\ref{eq_fp}) is the main result of the present work, from which one can immediately see that the very presence of DMI leads to an off-axial focusing of spin waves, i.e., $y_{f}\neq0$ if $D_{2}\neq0$, [see the red dot in Fig. \ref{fig2}(b)]. DMI can also modify the lateral shift $x_{f}$.

\section{Numerical results}\label{sec3}
To verify our theoretical results (\ref{eq_fp}), we perform full micromagnetic simulations using MuMax3 \cite{Vansteenkiste2014}. We consider a heterogeneous magnetic film with length $800$ nm, width $400$ nm, and thickness $2$ nm. The radius of the semi-circular interface is $R=200$ nm. The magnetic anisotropy constant and the DMI strength are $K_{1}=1\times10^{6}$ $\mathrm{J/m^{3}}$ and $D_{1}=0$ $\mathrm{mJ/m^{2}}$ in the left region (white) and $K_{2}=2\times10^{5}$ $\mathrm{J/m^{3}}$ and $D_{2}=3$ $\mathrm{mJ/m^{2}}$ in the right region (gray), see Fig. \ref{fig1}. Other magnetic parameters are considered as homogeneous: $M_{s}=1\times10^{6}$ $\mathrm{A/m}$, $A=15$ $\mathrm{pJ/m}$, and $\alpha=1\times10^{-4}$. An example of such a material system is the asymmetric Pt$/$Co$/$Al$_{2}$O$_{3}$. The difference of magnetic anisotropy can be realized by electrical tuning \cite{Ruiz2013}. As an important benchmark for frequency, the spin-wave gaps are $(\omega_{K_{1}}+\omega_{m}/2)/2\pi=73.6$ GHz in the left domain and $[\omega_{K_{2}}+\omega_{m}/2-(D_{2}^{\ast})^{2}/(4A^{\ast})]/2\pi=20.4$ GHz in the right domain, respectively, in the absence of external magnetic field. Absorbing boundary conditions have been adopted to avoid the spin-wave reflection by film edges \cite{Venkat2018}.

Next, we apply a sinusoidal monochromatic microwave field $\mathbf{H}_{\mathrm{ext}}=h_{0}\sin(\omega t)\hat{z}$ in a narrow rectangular area [thick black line shown in Fig. \ref{fig3}(a)] to excite the spin waves. We set the amplitude and frequency of the oscillating field as $\mu_{0}h_{0}=10$ mT and $\omega/2\pi=75$ GHz, respectively. Numerical results obtained from micromagnetic simulation are shown in Fig. \ref{fig3}(a). We also calculate the spin-wave intensity based on the formula $I(x,y)=\int_{0}^{t}[\delta m_{y}(x,y,t)]^{2}dt$, as plotted in Fig. \ref{fig3}(b). In Figs. \ref{fig3}(a) and \ref{fig3}(b), one can observe an interference pattern in the left domain, which is caused by the spin-wave reflection by the semi-circular interface. In Fig. \ref{fig3}(b), we clearly find a spin-wave focusing emerging in the right domain. Notably, the focal point obtained from the numerical simulation agrees excellently with the analytical formula Eq. (\ref{eq_fp}) [red point shown in Fig. \ref{fig3}(b)]. In addition, we note that the wavelength of spin waves is significantly shrunk across the interface [see Fig. \ref{fig3}(a)], which is due to the expansion of the radius of the isofrequency circle, as shown in Fig. \ref{fig2}(a).
This phenomenon can be considered as an effective method to generate the short-wavelength spin waves, which were conventionally realized by an interface with thickness step \cite{Stigloher2016,Stigloher2018} or by using the magnetization precession in periodic ferromagnetic nanowires on the top of a neighboring magnetic film \cite{Liu2018}.
\begin{figure}
  \centering
  \includegraphics[width=0.52\textwidth]{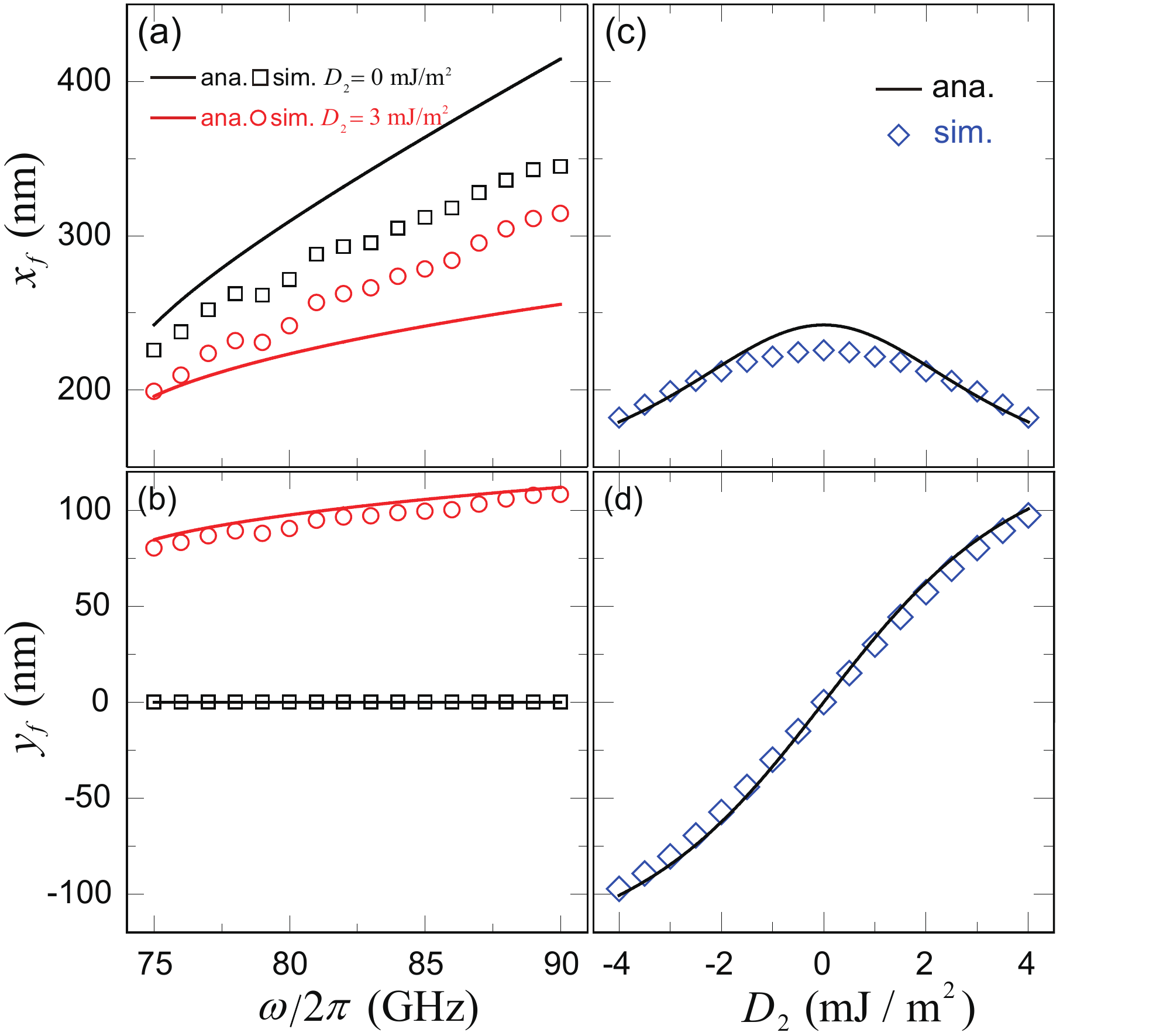}\\
  \caption{(a)-(b) Coordinate of the focal point ($x_{f},y_{f}$) as a function of the frequency $\omega$ for $D_{2}=0$ $\mathrm{mJ/m^{2}}$ (black squares) and 3 $\mathrm{mJ/m^{2}}$ (red circles). (c)-(d) The dependence of ($x_{f},y_{f}$) on the DMI constant $D_{2}$ for $\omega/2\pi=75$ GHz. Symbols are numerical data from micromagnetic simulations and solid curves represent the analytical formula (\ref{eq_fp}).}\label{fig4}
\end{figure}
\begin{figure}
  \centering
  \includegraphics[width=0.48\textwidth]{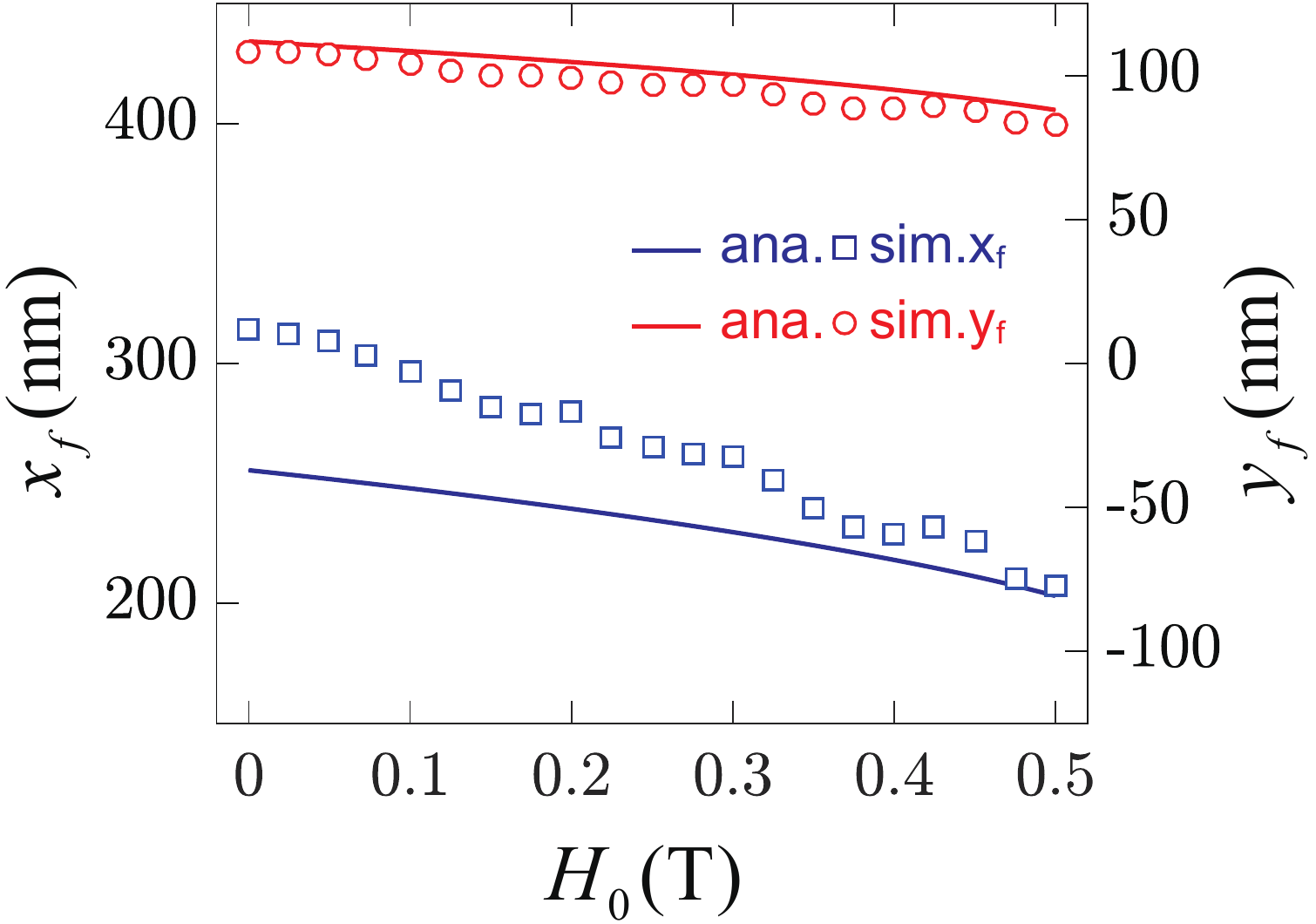}\\
  \caption{Dependence of ($x_f,y_f$) on the external field $H_0$ for $\omega /2\pi=90$ GHz. Blue squares and red circles correspond to numerical data from micromagnetic simulations. Solid curves represent the analytical formula (\ref{eq_fp}).}\label{fig5}
\end{figure}

Figures \ref{fig4}(a) and \ref{fig4}(b) show the coordinate of the focal point $(x_{f},y_{f})$ as a function of the frequency $\omega$ for two DMI constants. When $D_{2}=0$ $\mathrm{mJ/m^{2}}$, $x_{f}$ increases with the spin-wave frequency $\omega$ and $y_{f}$ is exactly zero. It recovers the expectation that the focal point should be on the lens axis ($y=0$) in non-chiral magnetic media. In the presence of DMI ($D_{2}=3$ $\mathrm{mJ/m^{2}}$), both $x_{f}$ and $y_{f}$ increase with $\omega$, which demonstrates an off-axial focusing of spin waves. In Figs. \ref{fig4}(c) and \ref{fig4}(d), we plot the dependence of the coordinate of the focal point on the DMI constant for $\omega/2\pi=75$ GHz. One can see that $x_{f}$ decreases and $y_{f}$ increases with the increasing of $|D_{2}|$, respectively. Furthermore, the sign change of $D_{2}$ has no effect on $x_{f}$, but it reverses the sign of $y_{f}$. It can be understood from Eq. (\ref{eq_tht0}) that the sign change of $D_2$ can switch the sign of $\theta_{0}$, and thus leads to the sign change of $y_{f}$ according to Eq. (\ref{eq_fp}). In addition, we note that the magnitude of $x_{f}$ is always suppressed by DMI. Micromagnetic simulations agree very well with the analytical formula (\ref{eq_fp}) for the lateral shift $y_{f}$ but not for the horizontal one $x_{f}$. There are several reasons that account for this discrepancy. First, the paraxial approximation is adopted in the analytical model, while the incident angles of spin waves in micromagnetic simulation are distributed broadly ($-90^{\circ}<\theta_{i}<90^{\circ}$). Second, the saturated magnetization is assumed in our theory. However, the spin canting cannot be avoided at the DMI interface (not shown) \cite{Wang2019}, which would change the spin-wave propagation and affect the focal point. Lastly, the coordinate of the focal point is obtained numerically by weighted averaging based on the spin-wave intensity, which might lead to sizeable errors.

In the above simulations, we have assumed a vanishing external magnetic field, while it could be an effective knob to manipulate the focusing of spin-wave lens. Figure \ref{fig5} plots the dependence of $(x_f,y_f)$ on the external magnetic field $H_0$. In the calculations, we set the DMI constant $D_2=3$ mJ/m$^2$ and the spin-wave frequency $\omega/2\pi=90$ GHz. It shows that the horizontal shift $x_f$ can be strongly suppressed by the magnetic field while the lateral shift $y_f$ is less sensitive to it. Numerical results (symbols) are consistent with theoretical formula (curves).

\section{Conclusion}\label{sec4}
In summary, we have theoretically investigated the DMI-induced off-axial focusing of a spin-wave lens. The generalized magnonic Snell's law at the semi-circular interface and the coordinate of the focal point were analytically derived. We showed that the induced lateral (horizontal) focal-point shift is an odd (even) function of the DMI parameter. While the shifts in both directions increase with the increasing of spin-wave frequency, they are suppressed by the external magnetic field. Full micromagnetic simulations were implemented to compare with theoretical predictions with good agreement. Our findings will help to understand the DMI effect on the spin-wave propagation across curved interface and to design chiral spin-wave optical elements for practical magnonic devices in the future.
\section{Acknowledgment}
This work was supported by the National Natural Science Foundation of China (Grants No. 11604041 and 11704060), the National Key Research Development Program under Contract No. 2016YFA0300801, and the National Thousand-Young-Talent Program of China. Z.W. acknowledges financial support from the China Postdoctoral Science Foundation under Grant No. 2019M653063.

\end{document}